\documentclass[12pt]{article}
\usepackage{epsf,amsfonts,amssymb,epsfig}
\newcommand{\bref}[1]{(\ref{#1})}

\newcommand{\be}{\begin{equation}}
\newcommand{\ee}{\end{equation}}
\newcommand{\ben}{\begin{displaymath}}
\newcommand{\een}{\end{displaymath}}
\newcommand{\bea}{\begin{eqnarray}}
\newcommand{\eea}{\end{eqnarray}}
\newcommand{\bean}{\begin{eqnarray*}}
\newcommand{\eean}{\end{eqnarray*}}
\newcommand{\nn}{\nonumber \\}
\newcommand{\ba}{\begin{array}}
\newcommand{\ea}{\end{array}}
\newcommand{\bi}{\begin{itemize}}
\newcommand{\ei}{\end{itemize}}

\newcommand{\str}{\mbox{str}}

\def\a{\alpha}

\def\b{\beta}
\def\g{\gamma}
\def\bg{\bar{\gamma}}
\def\G{\Gamma}

\def\s{\sigma}
\def\e{\epsilon}

\def\vp{\varphi}

\def\otaula{\begin{tabular}}
\def\ctaula{\end{tabular}}

\renewcommand{\t}{\theta}

\def\bnum{\begin{enumerate}}
\def\enum{\end{enumerate}}

\def\CR{\mathbb{R}}
\def\CM{\mathcal{M}}

\def\espai{\;\;\;\;\;\;\;}

\def\s{\sigma}
\def\8M{$\CM_8$}
\def\k{\kappa}
\def\be{\begin{equation}}
\def\ee{\end{equation}}
\def\G{\Gamma}
\def\g{\gamma}
\def\ei{e^{\underline{i}}}

\def\e1{e^{\underline{1}}}
\def\1u{\underline{1}}
\def\2u{\underline{2}}

\def\0u{\underline{0}}
\def\e{\epsilon}
\def\target{$\CR^{1,1}\times \mathcal{M}_8$ }
\def\target2{$\CR^{1,1}\times \mathcal{M}_8$,}
\def\9G{\G_{\underline{9}}}
\def\yvec{\vec{y}}
\def\rvec{\vec{r}}

\def\a{\alpha}
\def\b{\beta}
\def\undos{{1\over 2}}

\def\we{\wedge}

\def\ads{$AdS_5 \times S^5$ }

\newcommand{\caln}{\mbox{${\cal N}$}}
\newcommand{\calo}{\mbox{${\cal O}$}}

\newcommand{\pa}{\partial}

\newcommand{\sac}{\, , \qquad}

\def\w{\omega}
\def\hg{\hat{\gamma}}

\newcommand{\sla}[1]{\slash \!\!\!\! #1}

\def\ex{i \tau_2}

\begin{document}

\begin{titlepage}

\vfill

\begin{flushright}
hep-th/0505243\\
\end{flushright}
\vfill

\begin{center}

\baselineskip=16pt

{\Large\bf Marginal deformations of \caln=4 SYM \\*[5pt]
  and Penrose limits with continuum spectrum}

\vskip 1.3cm

Toni Mateos

\vskip 1cm
{\small{\it
Blackett Laboratory, Imperial College\\ Prince Consort Rd\\London, SW7 2AZ, U.K\\}}

\vskip .2cm

mateos@imperial.ac.uk

\end{center}

\vfill

\begin{center}
\textbf{Abstract}
\end{center}

\begin{quote}
We study the Penrose limit about a null geodesic with 3 equal
angular momenta in the recently obtained type IIB solution dual
to an exactly marginal $\gamma$-deformation of $\caln=4$ SYM.
The resulting background has non-trivial NS 3-form flux
as well as RR 5- and 3-form fluxes.
We quantise the light-cone Green-Schwarz action and show
that it exhibits a continuum spectrum. We show that this is
related to the dynamics of a charged particle moving in a Landau plane
with an extra interaction induced by the deformation.
We interpret the results in the dual $\caln=1$ SCFT.

\end{quote}

\vfill

\end{titlepage}

\tableofcontents

\vskip 2cm

\section{Introduction}
\label{sec-introduction}

Field theories with conformal symmetry often correspond to isolated points in the space of
couplings. The reason is that conformal symmetry implies the vanishing of all $\beta$-functions,
which imposes one relation for each coupling. As is well known~\cite{Leigh:1995ep},
supersymmetric field theories can circumvent
this argument because the $\beta$-functions can be expressed in terms of the anomalous dimensions
of the fundamental chiral fields. If the theory has less anomalous dimensions than couplings, then
not all the relations that follow from the vanishing of the $\beta$-functions are independent,
leading to a continuous family of conformal theories.
If, in addition, these field theories have a supergravity dual, the AdS/CFT correspondence~\cite{M,W,GKP} implies
the existence of continuous families of supergravity solutions with $AdS$ factors.

Lunin and Maldacena~\cite{LM} have recently provided a method to construct such families for
field theories with, at least, $U(1)_1 \times U(1)_2$ global symmetry. The essence of the method
is to use an $SL(2,\CR) \subset SL(3,\CR)$ transformation of the full $SL(3,\CR) \times SL(2,\CR)$ duality
group of IIB supergravity compactified along the corresponding $U(1)_1\times U(1)_2$ torus.
Note that exactly the same method had been used a few years ago
to generate the holographic duals of noncommutative field theories~\cite{MR,Hashimoto:1999ut}, the only difference
being that the torus is now transverse to the branes.

In this paper we will concentrate on the exactly marginal deformations of the 4d \caln=4 $SU(N)$ SYM which
are generated by this procedure.
The mentioned resemblance with noncommutative deformations
led the authors of~\cite{LM} to propose that the resulting field theories are such that
the standard products among the fields are replaced by
\be \label{nc-product}
\Phi_1 \, \Phi_2 \, \rightarrow \, e^{i \pi \g(Q^1_{\Phi_1} Q^2_{\Phi_2}-Q^1_{\Phi_2}Q^2_{\Phi_1})} \Phi_1 \Phi_2 \,,
\ee
where $(Q^1_{\Phi_i},Q^2_{\Phi_i})$ are the charges of $\Phi_i$ under the $U(1)_1 \times U(1)_2$ action,
and $\g$ is the deformation parameter.\footnote{In general, $\g$ can be complex, in which
case the letter $\beta$ is often used instead of $\g$. We will only consider the case when $\g$ is {\it real},
and consequently use the name $\g$-deformation.}$^,$\footnote{See also
\cite{Frolov:2005ty,Frolov:2005dj,Beisert:2005if,Benvenuti:2005cz,Benvenuti:2005ja}
for recent work on this theory, and \cite{Ahn:2005vc,Gauntlett:2005jb} for extensions to superconformal
field theories in three dimensions.}

Let us stress that this modification does not lead to a spacetime noncommutative theory, it just
introduces some phases in the operators of the theory; for example, the \caln=4 superpotential is modified
\be
tr \left( \Phi_1\Phi_2\Phi_3  -\Phi_1\Phi_3\Phi_2 \right) \, \rightarrow \,
tr \left(e^{i\pi \g}  \Phi_1\Phi_2\Phi_3   - e^{-i\pi \g} \Phi_1\Phi_3\Phi_2 \right) \,,
\ee
where the three $\Phi_i$ are the complex $\caln=1$ chiral superfields, each with unit charge with respect
to a generator $J_{\phi_i}$ of the Cartan subalgebra of the $SO(6)$ R-symmetry.
This is precisely one of the deformations of $\caln=4$ that had been proven to be exactly marginal
purely from field theoretical arguments~\cite{Leigh:1995ep}. In particular, it breaks $\caln=4$ to \caln=1 while
preserving only the Cartan subgroup of $SO(6)$.

Among the many new features of the deformed theory, we will be concerned with the fact that the set of chiral
operators is modified.
Recall that in the undeformed theory there exist 1/2-BPS operators with any
set of integer values of $(J_{\phi_1} , J_{\phi_1},J_{\phi_1})$, given simply
by
\be
\label{half-bps}
\calo = \str (\Phi_1^{J_{\phi_1}} \Phi_2^{J_{\phi_2}} \Phi_3^{J_{\phi_3}}) \,,
\ee
where 'str' stands for 'symmetrized trace'. Once the deformation is turned on, these operators
are multiplied as in~\bref{nc-product}, introducing a complicated set of relative phases
among the various terms in the sum. The result is that for generic
values of $\g$ the spectrum of single-trace 1/2-BPS operators reduces to those carrying
either of the following charges~\cite{LM,Berenstein:2000hy,Berenstein:2000ux}
\be \label{charges}
(J_{\phi_1} , J_{\phi_2},J_{\phi_3})\,\,=\,\,(n,0,0)\,\,,\,\,(0,n,0)\,\,,\,\,(0,0,n)\,\,,\,\,(n,n,n)
\sac n \in \mathbb{Z} \,.
\ee
The supergravity counterpart of this fact is as follows. In the original $\caln=4$, the dual
background is \ads. All null geodesics lying inside the $S^5$ are $SO(6)$-related, so we can
start with one carrying only $J_{\phi_1}$ charge, apply $SO(6)$ rotations and generate others
where the three $J_{\phi_i}$ are non-zero. In the background dual to the deformed $\caln=1$ SCFT,
the metric on the $S^5$ has been deformed so that we only have a $U(1)^3$ isometry at our
disposal. Thus geodesics carrying charges as in~\bref{charges} are isometrically
inequivalent.
Indeed, the background preserves an extra $\mathbb{Z}_3$ symmetry which
permutes the first three states of~\bref{charges} and leaves the fourth one invariant.
As a result, there are essentially two inequivalent Penrose limits about BPS geodesics
in the deformed geometry.

\subsection{Results and spectrum}
\label{sec-results}

In this paper we will study how some properties of the deformed $\caln=1$ SCFT
are realised in the string theory dual. With this aim in mind, we will take
the Penrose limit about the $(n,n,n)$ geodesic.\footnote{We refer the reader to~\cite{NP}
for a study of string theory in the Penrose limit about the $(n,0,0)$ geodesics, noting that these authors
remarkably discovered the corresponding pp-wave geometry even before~\cite{LM} appeared.}
We first show that the resulting background preserves 20 of the 32 supersymmetries.
We then show that,
despite the fact that such background (see eq.\ref{ours}) has three different types of fluxes turned on,
namely the NS 3-form and the RR 5- and 3-forms, it turns out that the
Green-Schwarz action is quadratic in the light-cone gauge, and therefore
exactly solvable.

We will show that the Penrose limit focuses on operators with
\be \label{ourpenrose}
\Delta, J_{\phi_1} ,J_{\phi_2}, J_{\phi_3} \, \sim \, \calo(R^2) \sac \Delta -
(J_{\phi_1} +J_{\phi_2}+ J_{\phi_3}) \, = \, \calo(1) \,,
\ee
and forces us to scale $\g$ such that
\be \label{gammascaling}
\hg \, \equiv \, {\g  \over \sqrt{3}} {R^2 \over l_s^2} \, = \, \mbox{fixed} \,,
\ee
where $R^4/l_s^4 \sim g_s N  \rightarrow \infty$ in the limit. In particular, \bref{gammascaling}
requires $\g\ll 1$, ensuring that string theory in the resulting
background is not $SL(2,\mathbb{Z})$ equivalent to string theory in the maximally
supersymmetric pp-wave.

Before writing down the string theory spectrum in the Penrose limit of the deformed
background, let us first discuss what the results would be in the equivalent
Penrose limit (focusing on states with quantum numbers as in~\bref{ourpenrose}) of $AdS_5 \times S^5$.
When $\gamma=0$ there are a huge number of operators scaling as required by \bref{ourpenrose}
and with $\Delta - (J_{\phi_1} +J_{\phi_2}+ J_{\phi_3})=0$.
These are obtained from the operator~\bref{half-bps} by removing $\Phi_i$ and inserting one $\Phi_j$,
with $i\neq j$.
We will refer to this as {\it exchanges} of $\Phi_i \leftrightarrow \Phi_j$.
For large $J$'s, this procedure generates infinitely many new states.
The quantization of the string $\s$-model in corresponding Penrose limit of
reflects this fact by producing a ground state with an {\it infinite discrete} degeneracy.
Some of the particle-like modes (those which are homogeneous along the string) have dynamics
corresponding to that of a particle in a plane threatened by a constant magnetic field,
a Landau problem, which explains the ground state infinite degeneracy~\cite{NP}.

It is interesting to see how string theory knows about the fact that, as soon as $\gamma\neq 0$,
none of the previous exchanges  produces BPS-protected operators any longer, as we read from the
series~\bref{charges}. We would therefore expect the string theory ground state to be unique, and
we expect that all mentioned exchanges lead to operators with $\Delta-J >0$. More explicitly,
we expect the latter operators to have  
$\Delta-(J_{\phi_1} +J_{\phi_2}+ J_{\phi_3}) = f(\gamma,J_{\phi_i})$, with $f$ a positive function that vanishes if $\g=0$ or
if all $J_{\phi_i}$ are equal; it should therefore be proportional to $\gamma$
and depend only on the differences between the $J_{\phi_i}$.

\begin{figure}[t] \begin{center}
\includegraphics[width=5 cm,height=3 cm]{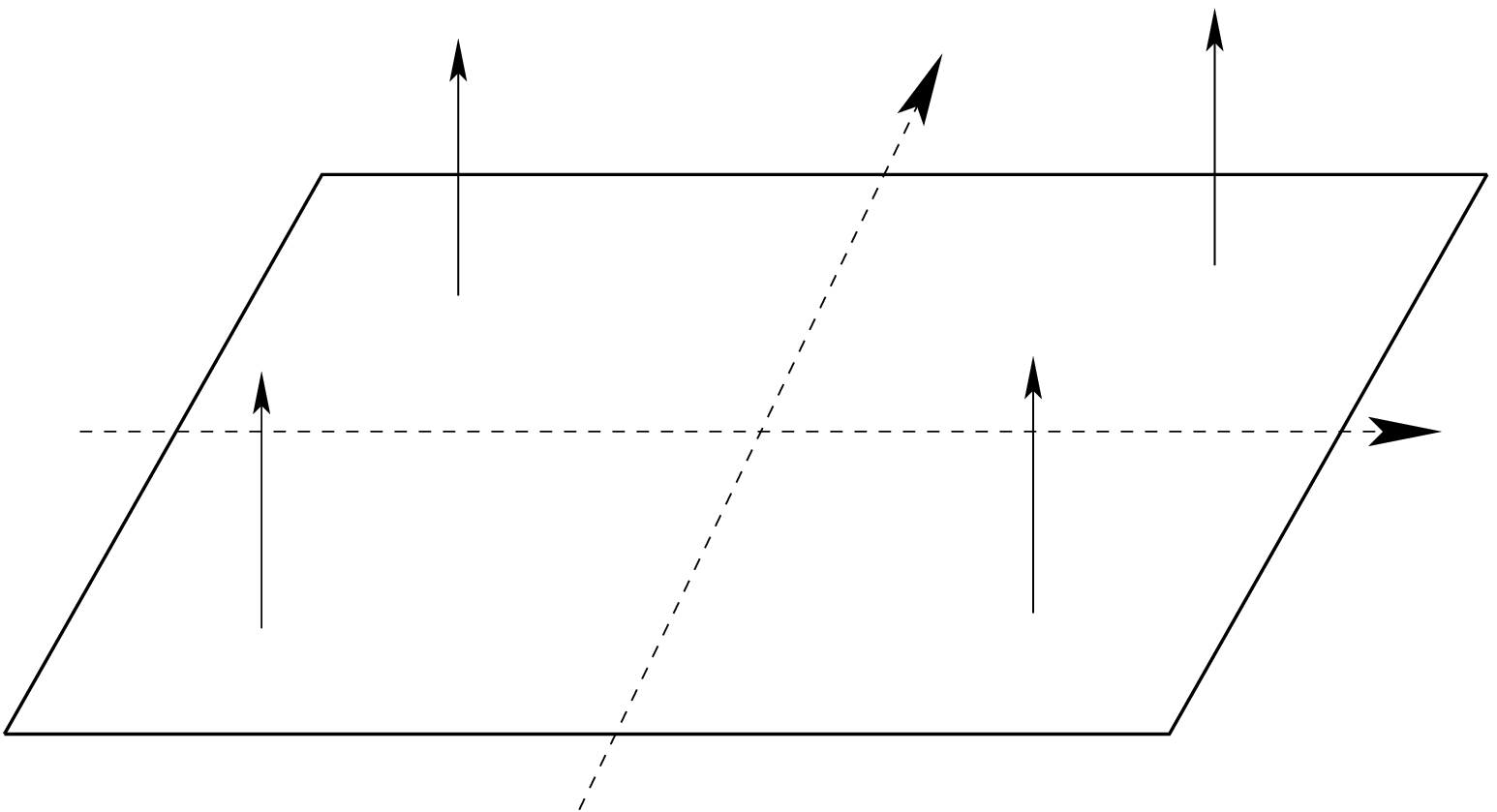} $\espai\espai\espai$
\includegraphics[width=6 cm,height=4 cm]{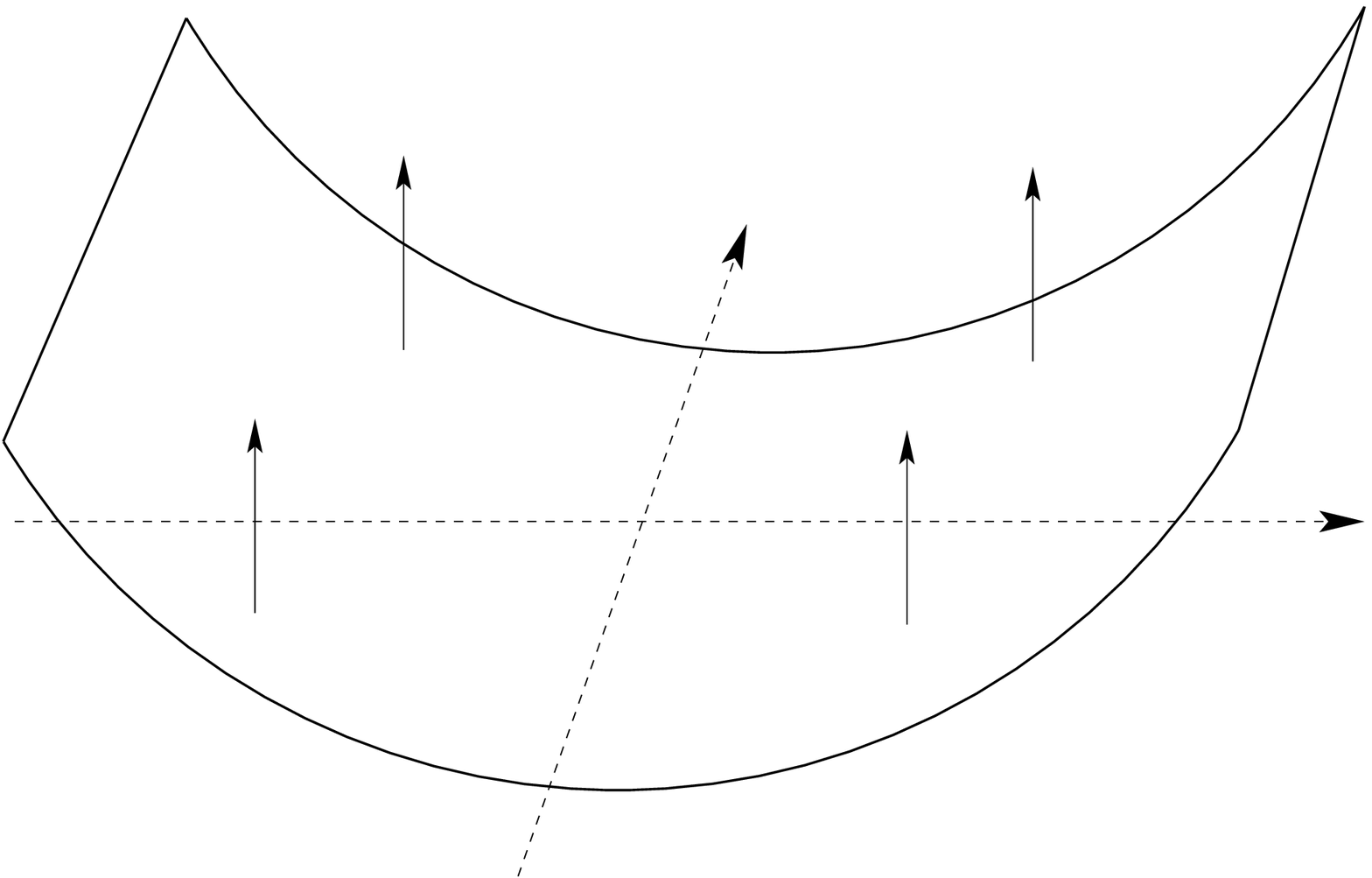}
\caption{The physics as seen by the particle excitations of the string. Vertical
arrows represent the magnetic field.
Left: a Landau problem. Right: the modified Landau problem when $\g\neq 0$. \label{fig}}
\end{center}
\end{figure}

At the level of the string zero modes,
we find that the turning on of $\gamma$ introduces an extra interaction in the Landau problem.
This interaction is an attractive quadratic potential along only one of the axis of the plane.
We can visualize the problem as imagining that the plane is bent as in figure \ref{fig}.
Note that the asymptotics of the potential are radically changed no matter how small $\g$ is.
We find that this leads to a very different behavior of the system: the infinite
discrete degeneracy is lifted, leaving behind a unique ground state and a continuous
free-particle spectrum. Explicitly, we find the light-cone energy of these modes is
\be
E_{l.c.} = \mbox{const.} \times \g^2 \times  p^2 \,,
\ee
where $p$ is the momentum along the undeformed direction of the Landau plane in fig.\ref{fig}.
We will have more to say about the field theory interpretation of these results in section \ref{sec-discussion}.

Let us finally gather here the results concerning the spectrum of the string, which is derived in
section \ref{sec-quant}. Note that the $SL(2,\CR)$ deformation leaves the $AdS_5$ part intact;
as a result, the Penrose limit produces 4 transverse directions which are identical to
the maximally supersymmetric pp-wave limit of~\cite{BMN}. However, the $S^5$ part of the geometry
is modified and leads to 4 'modified Landau directions' with more complicated interactions.
We will show that the background preserves 20 supersymmetries, 4 of them being 'supernumerary'~\cite{Cvetic:2002nh}
and therefore linearly realised. So, for simplicity, we only include the
complete bosonic spectrum in the following table.

\small
\label{taula}
\begin{center}
\begin{tabular}{|c|c|ccc|c|}
\hline
 & level  & &number of modes   & &$\Delta - (J_{\phi_1}+J_{\phi_2}+J_{\phi_3})$ \cr \hline
four pp-wave directions  & $n=0$    & & 4   & & 1   \cr
(origin: $AdS$)          & $n\neq 0$ && 4 left-mov + 4 right-mov &  & $\sqrt{1+{n^2 \over (\mu \a' p_v)^2}}$  \cr \hline
four modified   &   $n=0$ && 2 && continuous \cr
Landau directions &   $n=0$ && 2 & &2 \cr
(origin: $S^5$)       & $n \neq 0$ && 4 left-mov + 4 right-mov & &$\pm 1 + \sqrt{1+{n^2 \over (\mu \a' p_v)^2}}$ \cr \hline
\end{tabular}
\end{center}
\normalsize
{\vskip .5cm}

Note that despite the fact that the background (and therefore the whole GS action) depends
in a very non-trivial way on the deformation parameter $\g$, most modes
have energies which are independent of it.
The factors $\pm 1$ of the Landau frequencies relative to the standard pp-wave
ones are easily explained~\cite{JO,NP} by the twisting in the worldsheet hamiltonian
that is induced when changing from coordinates adapted to focus on geodesics
with charges $(n,0,0)$ to charges $(n,n,n)$.

The organization of the paper is as follows. In section \ref{sec-penrose} we discuss the BPS null geodesics
in the $\g$-deformed background and perform the Penrose limit along one of them. Section \ref{sec-susy}
shows that the resulting background preserves 20 supersymmetries. Section \ref{sec-quant} is devoted
to the quantization of the bosonic and fermionic sectors of the GS action. Finally,
we discuss the implications of our results in section \ref{sec-discussion}.

\section{The Penrose limit of the deformed background}
\label{sec-penrose}

In this section we comment on the different null geodesics that the $\g$-deformed background
has and we perform the Penrose limit that focuses on one of them.

We start with the type IIB \ads background, with metric, RR 5-form and dilaton given by
\bea
\label{ads5}
ds^2 &=& R^2 \, \left( -\cosh^2 \rho dt^2 + d\rho^2 +\sinh^2 \rho d\Omega_3^2 \right)
+ R^2 \sum_{i=1}^3 \left( d\mu_i^2 + \mu_i^2 d\phi_i^2 \right)\,, \nn
F_5 &=& 4 R^4 e^{-\phi_0} \left( \w_{AdS_5} + \w_{S^5} \right) \,, \nn
e^{\phi} &=& e^{2\phi_0} \,,
\eea
with $(\mu_1,\mu_2,\mu_3)=(\cos \a,\sin\a \cos\t,\sin\a \sin\t)$ and $R^4=4\pi e^{\phi_0}N l_s^4$.
Out of the $SO(6)$ isometry group of the round $S^5$, the coordinates
chosen in~\bref{ads5} exhibit only a manifest $U(1)^3$ subgroup which acts as shifts
of the 3 angles $\phi_i$. These will lead to conserved charges in the string worldsheet
which we will name $J_{\phi_i}$. In the dual gauge theory, each of these generators rotates
the corresponding complex $\caln=1$ scalar superfield $\Phi_i$ with unit charge.

To obtain the background dual to the $\b$-deformed SCFT, we first need to select
two $U(1)$ symmetries, compactify along them to 8d, and then perform an $SL(2,\CR)$
transformation. Lunin and Maldacena~\cite{LM} chose a $U(1)^2$ subgroup generated by a certain
linear combination of the $J_{\phi_i}$ defined above. Namely, by defining
three new angles $(\psi,\vp_1,\vp_2)$ via
\be
\phi_1 =  \psi+\vp_1+\vp_2 \sac \phi_2 = \psi-\vp_1 \sac \phi_3 = \psi-\vp_2  \,,
\ee
the chosen $U(1)^2$ acts as shifts of $\vp_1$ and $\vp_2$. For future reference, we
will label the charges corresponding to shifts of the new 3 angles by $(J_{\psi},J_{\vp_1},J_{\vp_2})$,
related to the previous ones by
\bea \label{angular}
J_{\psi}=J_{\phi_1}+J_{\phi_2} +J_{\phi_3} \sac J_{\vp_1} = J_{\phi_1}-J_{\phi_2}
\sac J_{\vp_2}=J_{\phi_1}-J_{\phi_3} \,.
\eea
Note that all the scalars $\Phi_i$ have charge 1 under $J_{\psi}$ and that the $J_{\vp}$'s measure
deviations from $J_{\Phi_1}=J_{\Phi_2}=J_{\Phi_3}$.
Being the generators of rotations along the torus, $(J_{\vp_1},J_{\vp_2})$ are to be
identified with the charges $(Q^1,Q^2)$ of equation~\bref{nc-product}.

The resulting $\g$-deformed background has, in addition to the previous IIB fields, nontrivial
RR and NS 3-forms. In order to analyze its null geodesics it will suffice by now to concentrate
only on the resulting metric,
\bea \label{deformedads}
ds^2_{\g} &=& R^2 \left[ ds^2_{AdS_5}+ \sum_{i=1}^3 \left( d\mu_i^2 + G \mu_i^2 d\phi_i^2 \right)
+{R^4\over l_s^4} \g^2 G \mu_1^2\mu_2^2\mu_3^2 (\sum_{i=1}^3 d\phi_i )^2 \right] \,, \nn
G^{-1}&\equiv& 1+{R^4\over l_s^4} \g^2 (\mu_1^2 \mu_2^2 + \mu_2^2 \mu_3^2 +\mu_1^2 \mu_3^2 ) \,.
\eea
All other fields are written in the appendix~\ref{sec-appendix-full}.
A quick look at the metric reveals that the $S^5$ has been deformed in such a way that
the original $SO(6)$ isometry has been broken to $U(1)^3$, out of which only the $U(1)$
generated by $J_{\psi}$ remains an $R$-symmetry in the dual $\caln=1$ SCFT.
Let us label by $S^5_{\g}$ the deformed sphere with metric as in~\bref{deformedads}
and note that because the deformation is a continuous one,
its topology is still that of an $S^5$.

We now wish to perform the Penrose limit about a null geodesic lying inside the $S^5_\g$.
In the original round $S^5$, we could use the full $SO(6)$ symmetry to relate all such
geodesics. In contrast, we now only have a $U(1)^3$ group at our disposal and, as a result,
Penrose limits about different geodesics may give rise to non diffeomorphic metrics.
We do not aim at studying all such geodesics here, but we will rather concentrate on those
that are BPS. At this point, we use information from the gauge theory~\cite{LM,Berenstein:2000hy,Berenstein:2000ux},
where we know that,
for generic $\g$, the spectrum of single-trace BPS operators reduces to those carrying
either of the following charges
\be \label{charges2}
(J_{\phi_1} , J_{\phi_2},J_{\phi_3})\,\,=\,\,(n,0,0)\,\,,\,\,(0,n,0)\,\,,\,\,(0,0,n)\,\,,\,\,(n,n,n)
\sac n \in \mathbb{Z} \,.
\ee
The corresponding null geodesics can be parametrized as follows:
using $\tau$ as the worldline coordinate, and setting $t=\phi_1=\phi_2=\phi_3=\tau$,
the charges in~\bref{charges2} are carried by massless particles along
geodesics with
\be
(\mu_1^2,\mu_2^2,\mu_3^2) \,\,= \,\,(1,0,0)\,\,,\,\,(0,1,0)\,\,,\,\,(0,0,1)\,\,,\,\,
({1/ 3},{1/ 3},{1/ 3}) \,,
\ee
respectively. From the $AdS_5$ perspective they are just a point sitting
at the origin in global coordinates.

The quantization of the string sigma model in the Penrose limit of the first
three geodesics was studied by Niarchos and Prezas~\cite{NP}. Here we will instead consider
the fourth case. Such Penrose limit can be obtained by defining
\bea
t&=&u \sac  \rho={r \over R} \sac \psi=u+{v\over R^2} \sac \a = \a_0 + {y^1 \over R}
\sac \t = {\pi \over 4} + \sqrt{3\over 2} {y^3 \over R} \,,\nn
\vp_1 &=& \left(1\over 2 G\right)^{1/2} {-y^2 +\sqrt{3} y^4 \over R} \sac
\vp_2 = -\left(1\over 2 G\right)^{1/2} {y^2 +\sqrt{3} y^4 \over R} \,,
\label{rescalings}
\eea
with $\a_0=\cos^{-1} 1/\sqrt{3}$,
and then sending $R\rightarrow \infty$ while keeping $\hg={\g \over \sqrt{3}} {R^2 \over l_s^2}$ fixed.
In the limit, $G$ tends to a constant,
\be
G \rightarrow {1 \over 1 +\hg^2} \,,
\ee
and this is the value for $G$ that will be used in the rest of the paper.

Using the expressions for the rest of the fields of the $\g$-deformed solution~\bref{allofit},
we find that the resulting type IIB configuration is
\bea
ds^2_{IIB}&=&2 du dv -[r^2 + 4 G\hg^2 \mu^2 ((y^1)^2+(y^3)^2)]du^2 \nn && 
\espai\,\,\,\,\,\,               + 4 G^{1/2}\mu du (y^1 dy^2+y^3 dy^4) + d \rvec^2 +  d\yvec^2 \,,\nn
H_3 &=& 2 \hg G^{1/2} \mu \, (-dy^1 \we dy^4+dy^3 \we dy^2)\we du \,, \nn
F_3 &=&  -\hg 4 \mu \,\, du \we dy^1 \we dy^3 \,, \nn
F_5 &=&  4 \, \mu \, du \,\we \, ( dr^1 \we dr^2 \we dr^3 \we dr^4 +dy^1 \we dy^2 \we dy^3 \we dy^4) \,, \nn
e^{2 \phi} &= & e^{2 \phi_0} G \,,
\label{ours}
\eea
with $G^{-1}=1+\hg^2$. Here $\rvec$ and $\yvec$ parametrise two copies of $\CR^4$, with $r$ and
$y$ being the corresponding radial coordinates. We have introduced
a mass parameter $\mu$ via $(u,v) \rightarrow (\mu u,v/\mu)$; this ensures that the metric
tensor is dimensionless whereas all coordinates have dimensions of length.
Note that in string theory, the light-cone momenta 
correspond to
\be \label{string-charges}
{p_u\over \mu} = -i \partial_u = \Delta - (J_{\Phi_1} +J_{\Phi_2} +J_{\Phi_3} ) \sac
\mu p_v = -i \partial_v = R^{-2} (J_{\Phi_1} +J_{\Phi_2} +J_{\Phi_3} ) \,,
\ee
so that requiring that they be finite in the limit leads to~\bref{ourpenrose}.

Let us make a few remarks about the resulting background.
\begin{itemize}
\item Despite the fact that the dilaton is non-constant before taking the limit (see~\bref{allofit}), it becomes
constant (although different from the \ads case) after the limit.

\item A crucial point is that the whole background admits a covariantly constant null Killing vector $\partial_v$.
It can then be argued~\cite{Metsaev:2001bj,Metsaev:2002re,Russo:2002qj,Mizoguchi:2002qy} that the fermionic GS action~\cite{Green:1983wt}
is obtained from the flat one via the replacement of partial derivatives by the type IIB supercovariant ones
(appearing e.g. in the variation of the gravitino).

\item The metric exhibits a natural split between the four directions parametrized by $\rvec$
with $AdS$ origin, and the four parametrized by $\yvec$ with $S^5_\g$ origin;
whereas the former remain identical to those of the maximally supersymmetric pp-wave,
the latter will lead to more sophisticated physics.

\end{itemize}

We will find it more convenient to work in a slightly different gauge.
After the coordinate transformation $v \rightarrow v-G^{1/2}(y^1 y^2+y^3 y^4)$, the metric
and NS 3-form can be written as
\bea
\label{simpler}
ds^2_{IIB}&=&2 du dv -(r^2 + k_{i} y^i y^i) du^2 + 2  f_{ij} y^i dy^j du  +  d \rvec^2 +  d\yvec^2 \,,\nn
H_3 &=& h_{ij} dy^i \we dy^j \we du \,,
\eea
with $f_{ij}$ and $h_{ij}$ antisymmetric and only nonzero components
\be
\label{components}
f_{12}=f_{34}=\mu G^{1/2} \sac h_{14}=h_{23}=-\mu G^{1/2}\hg \sac k_{1}=k_{3}=4\mu^2 G \hg^2 \,.
\ee
This more compact notation will simplify the expressions throughout the paper.
It also helps identifying that the metric and the NS 3-form fall into a particular
case of the pp-waves studied in~\cite{Blau:2003rt}. It was shown there that despite the fact that
the crossed $dy^i du$ terms in the metric can in general be brought to $du^2$ terms
via coordinate transformations, the resulting coefficients are $u$-dependent unless the matrices
$f_{ij}$ and $k_{ij}=k_i \delta_{ij}$ commute. From~\bref{components} it is straightforward to check that this
is not so in our case, and we will therefore stick to the form~\bref{simpler} where
all coefficients are constant.

\section{Number of supersymmetries}
\label{sec-susy}

Let us find the number of supersymmetries that the bosonic type IIB configuration~\bref{ours}-\bref{simpler}
preserves. It is possible to show that the total number must be 20 without having to actually
solve the Killing spinor equations for the resulting background. The shortcut is based on the
observation made it~\cite{LM} that the Penrose limit and the $SL(2,\CR)$ transformation
used to deform \ads commute. In other words, our background~\bref{ours} can also be
obtained\footnote{We have checked explicitly that this alternative method leads to the same background.} as follows:
\begin{enumerate}
\item Take the Penrose limit about the $J_{\phi_1}=J_{\phi_2}=J_{\phi_3}$ geodesic of the
\ads metric~\bref{ads5}. This leads to the maximally supersymmetric pp-wave in 'magnetic'
coordinates\footnote{The name refers to the fact that the light-cone dynamics of a relativistic particle
in these coordinates is equivalent to that of a non-relativistic one in a magnetic field.}
\bea \label{magnetic}
ds^2_{IIB}&=&2 du dv -\mu^2 r^2 du^2 + 4 \mu  (y^1  dy^2 +y^3 dy^4) du  +  d \rvec^2 +  d\yvec^2 \,, \nn
F_5 &=&  4 \, \mu \, du \,\we \, ( dr^1 \we dr^2 \we dr^3 \we dr^4 +dy^1 \we dy^2 \we dy^3 \we dy^4) \,.
\eea
\item Compactify $y^2$ and $y^4$ on a torus,
perform the $SL(2,\CR)$ duality along this torus and, finally, decompactify $y^2$ and $y^4$.

\end{enumerate}

We can easily compute the explicit form of the 32 Killing spinors for~\bref{magnetic}. Only those
which happen to be independent of $(y^2,y^4)$ will survive the $SL(2,\CR)$ transformation.
The Killing spinor equations that follow from the variation of the IIB gravitino are
\bea \label{spinoreq}
 \left(\pa_M + {1\over 4} \w_M^{mn} \G_{mn} +{i \over 5!\,\, 16} (F_5)_{mnpqr}\G^{mnpqr} \G_M \right)\e
&=&0\,,
\eea
where $M=0,...,9$ is a curved spacetime index, $m,n..=0,...,9$ are tangent space ones, and $\e$
is a 10d positive chirality complex Weyl spinor. As explained in appendix~\ref{sec-appendix-conv}, we will use
the indices $\{i,j=1,..,4\}$ to denote the $y^i$ coordinates, and indices $\{a,b=5,...,8\}$ to denote the
$r^a$ ones.
We will use the standard pp-wave vielbeins
\be
e^v=dv - \mu^2 r^2 du  + 2\mu (y^1 dy^2+y^3 dy^4) \sac e^u=du \sac \e^a=dr^a \sac e^i=dy^i \,,
\ee
which brings the metric to $ds^2=2 e^u e^v + e^a e^a + e^i e^i$. See appendix~\ref{sec-appendix-conv}
for our index conventions. The non-vanishing components of the spin connection are
\be
\w^{ij}=-f_{ij} du \sac \w^{vi}=f_{ij} dy^j \sac \w^{va}=-\mu^2 r^a du \,,
\ee
with $f_{ij}$ being the same antisymmetric matrix as in~\bref{components} but with the deformation parameter
set to zero, i.e., $f_{12}=f_{34}=\mu$. Using the reasoning of~\cite{Blau:2001ne} as a guideline, we rewrite the Killing
spinor equation~\bref{spinoreq} as $(\pa_M + i \Omega_M) \e =0$ with
\bea
\Omega_v &=&0 \,, \label{eqv} \\
\Omega_u &=& - \undos f_{ij} \G_{ij} - \undos \mu^2 r^a \G_{va} +{\mu \over 4} \G_v (\G_{5678}
+\G_{1234} )\G_u \,, \label{equ}  \\
\Omega_{a} &=& {\mu \over 4} \G_v \G_{5678} \G_a \,, \label{eqa} \\
\Omega_{i} &=& {\mu \over 4} \G_v \G_{1234} \G_i + {i\over 2} f_{ij} \G_{vj} \label{eqi} \,.
\eea
The first immediate solutions are given by the 16 spinors subject to $\G_v \e =0$. For these, equations
(\ref{eqv},\ref{eqa},\ref{eqi}) imply that they can only depend on $u$; such dependence is fixed
by~\bref{equ}, which becomes a first-order linear ordinary differential equation with constant coefficients,
which has a unique solution for each initial value. Because all these 16 spinors are $(y^2,y^4)$-independent
they survive the $SL(2,\CR)$ transformation.

Let us now look at the form of the remaining 16 spinors. Nilpotency of $\G_v$ implies that
\be
\Omega_{a} \Omega_{b} =0 \sac  \Omega_{a} \Omega_{i} = 0 \sac  \Omega_{i} \Omega_{j} = 0
\sac \forall a,b,i,j \,.
\ee
This implies that both~\bref{eqa} and~\bref{eqi} are solved by spinors of the form
\be \label{extra16}
\e(u,r^a,y^i) = \left(1-i r^a \Omega_{a} -i y^i \Omega_{i} \right) \, \chi(u) \,.
\ee
It is straightforward to check that plugging this expression into the remaining equation~\bref{equ}
leads again to a  first-order linear ordinary differential equation with constant coefficients for
$\chi(u)$. The conclusion is that the remaining 16 Killing spinors are all of the form~\bref{extra16}.
Out of them, the only ones that do not depend on $(y^2,y^4)$ are those for which
\be
\Omega_{2} \chi =0 \sac \Omega_{4} \chi =0 \,.
\ee
Using the explicit values of the $\Omega$'s, these conditions can be more properly expressed as
\be \label{projection}
\G_{12} \chi = i \chi \sac \G_{34} \chi = i \chi \,,
\ee
each reducing the dimension of the space of solutions by 1/2. We conclude that only
4 of the 16 spinors~\bref{extra16} survive the $SL(2,\CR)$ deformation.
Physically, these projections select those spinors which are invariant under
rotations in the planes $(y^1,y^2)$ and $(y^3,y^4)$.

The conclusion is that the Penrose limit under consideration preserves $16+4=20$ supersymmetries.
The extra 4 ones are often called 'supernumerary'; a very important property~\cite{Cvetic:2002nh}  is that they
are the only ones that become linearly realised in the string worldsheet when the
light-cone gauge is imposed, leading to standard properties of supersymmetric field theories like
equality of bosonic and fermionic masses.

We end by noting that we have rederived the same results of this section by working out the Killing spinors
directly in the final background~\bref{ours}. We do not include the computations here, but just point out
that, in that case, both projections~\bref{projection} follow directly from the variation of the dilatino.

\section{Quantization of the string $\s$-model}
\label{sec-quant}

As explained in section \ref{sec-penrose}, the four directions parametrized by $\rvec$ are unchanged
with respect to the standard maximally supersymmetric pp-wave. We will therefore only
consider the four remaining ones, parametrized by $y^i$, with $i=1,2,3,4$.
We will work in the second coordinate system discussed in section \ref{sec-penrose}; therefore,
the configuration is as in~\bref{ours} but with the values of the metric and $B$-field written
in~\bref{simpler}.

A last remark is that we will not consider the contribution of the vacuum energy as this is guaranteed
to cancel among fermions and bosons due to our results of section~\ref{sec-susy}.

\subsection{Quantization of the bosonic sector}
\label{sec-quant-bos}

We start with the bosonic part of the closed string $\s$-model,
\be
S = -{1 \over 4 \pi \a'} \int d\tau d\sigma \, \left( \sqrt{-h} \, h^{\a\b} \, \pa_a X^M \pa_b X^N G_{MN}
+  \, \e^{\a\b} \, \pa_a X^M \pa_b X^N B_{MN} \right) \,,
\ee
with $0 \le \sigma \le \pi$. We choose to work with dimensionless worldsheet coordinates, so that the worldsheet
lagrangian and hamiltonian are dimensionless too; we will also work in conformal gauge $h_{\a\b}=\eta_{\a\b}$.
The equation of motion for the spacetime field $u$ allows us go to the light-cone,
\be
u=2 \a' p_v \, \tau  \equiv \kappa \tau \sac \kappa = 2 \a' p_v\,.
\ee
Because of this choice, the spacetime light-cone energy $E_{lc}$ is related to the worldsheet one $H_{lc}$ by
\be \label{rela}
E_{lc} = {H_{lc} \over 2 \a' p_v} \,.
\ee
Having fixed the gauge, the equations of motion for the remaining 4 spacetime fields $y^i$ which
originated from the $S^5_{\g}$ are
\bea
\label{eom}
-\ddot{y}^i + (y^i)'' +2 \kappa (f_{ij} \dot{y}^j - h_{ij} y'_j)- \kappa^2 k_i y^i &=&0 \,,
\eea
where, as usual, dots stand for $\partial_{\tau}$ and primes for $\partial_\s$.
To solve these equations, we expand the fields in Fourier modes
\be \label{field-exp}
y^i(\tau,\s) = \sum_{n=-\infty}^{\infty} y^i_n(\tau) e^{i 2n\s} \,,
\ee
which leads to
\be
\label{eom2}
-\ddot{y}^i_n +2 \kappa f_{ij} \dot{y}^j_n -(\kappa^2 k_i+4 n^2) y^i_n -4in \kappa h_{ij} y^j_n =0 \,.
\ee
Notice that the last is the only term that mixes the physics in the 12-plane with the physics in the 34-plane.
The solution to these equations for the particle-like modes ($n=0$) and the stringy modes $n\neq 0$
are qualitatively very different, and so we treat them separately.

\subsubsection{The stringy modes}
\label{sec-quant-1}

If $n\neq 0$, we can try to solve the equations by expanding each of the string modes in
a standard harmonic oscillator frequency ansatz, $y^i_n(\tau) \sim u^i(\w_n) e^{i w_n \tau}$.
Plugging this expression into~\bref{eom2} we obtain the following linear system of equations for the
vector $u^i(\w_n)$,
\be \label{eigen}
M_{ij}(w_n,n)u^j_n =0 \,,
\ee
with
\bea
M_{ij}(\w,n)&\equiv&(\w^2-\kappa^2 k_i -4 n^2)\delta_{ij}+ 2 i \kappa \w f_{ij}-4 i \kappa n h_{ij} \,.
\eea
This system admits non-trivial solutions only if the frequencies are such that
\bea
0= \det M_{ij}(\w_n,n) = \left[ (\w_n^2-4 n^2)^2-4 \kappa^2 \mu^2 \w_n^2  \right]^2 \,,
\eea
which leads to 4 different admissible frequencies, each with multiplicity 2. Because the determinant
does not depend on the sign of $\w_n$, it therefore suffices to give the explicit expression
for the 2 positive different roots, which are
\be \label{freqs}
\w_n^\pm = \pm \kappa \mu +\sqrt{\kappa^2 \mu^2 +4n^2} \,.
\ee
It is remarkable that, despite the fact that the equations~\bref{eom2}-\bref{eigen} depend
in a highly non-trivial way on $\hg$, the resulting frequencies are completely independent
of it.

Having found the 8 allowed frequencies we now need to determine the corresponding 8 null eigenvectors
$u^i (\w_n)$ satisfying~\bref{eigen}. It can be checked that all minors of the matrix $M_{ij}$
vanish, which implies that we cannot use the results of~\cite{Blau:2003rt}; we have to
diagonalise the system by brute force.
Because each frequency has multiplicity 2, we indeed
find a 2d degenerate vector space for any given $\w_n$, for which a convenient orthogonal
basis is provided by the following two vectors in $\CR^4$,
\bea \label{eigenfun}
u^i_1(\w_n)& =& (\w_n^2-4n^2\,,\,2i\w_n \kappa f\,\,,0\,,\,-4 i n h \kappa) \sac \nn
u^i_2(\w_n)& =& (0\,,\, 4 i n h \kappa\, ,\, \w_n^2-4n^2\,  ,\,2i\w_n \kappa f) \,.
\eea
Summarizing, the most general solution to~\bref{eom2} with $n \neq 0$ is a linear combination
of the 8 particular solutions described above,
\bea
\label{exp}
{y^i_n(\tau) \over \sqrt{2\a'}} &=&
\left[\xi_1^{(n)} \, u^i_1(\w_n^+) + \xi_2^{(n)} \,u^i_2(\w_n^+)\right] e^{i\w_n^+ \tau}
+\left[\xi_3^{(n)}\, u^i_1(\w_n^-) + \xi_4^{(n)}\, u^i_2(\w_n^-)\right] e^{i\w_n^- \tau} \nn
&+& \left[\xi_5^{(n)} \,u^i_1(-\w_n^+) + \xi_6^{(n)}\, u^i_2(-\w_n^+)\right] e^{-i\w_n^+ \tau}
+\left[\xi_7^{(n)}\, u^i_1(-\w_n^-) + \xi_8^{(n)}\, u^i_2(-\w_n^-)\right] e^{-i\w_n^- \tau}  \nonumber \,.
\eea
The arbitrary constant coefficients $(\xi_1,...,\xi_8)$ will become operators in the quantum theory.
Whereas the first 4 ones are associated to positive frequencies and, so, to left-moving excitations,
the last 4 ones are associated to negative frequencies and, so, to right-moving excitations.

Let us now proceed to the quantization of the system. The canonical momenta associated to the
dynamical fields $y^i(\tau,\s)$ are easily found to be
\be \label{can-mom}
\Pi_i(\tau,\s)  = \dot{y}^i(\tau,\s)\,\, -\,\, \kappa f_{ij} y^j (\tau,\s) \,.
\ee
Promoting the fields to operators and imposing the equal-time commutators
\bea
[ y^i(\tau,\s),\Pi_j(\tau,\s')] = i \delta^i_j \delta(\s-\s') \sac
[ y^i(\tau,\s),y^j(\tau,\s')]=0= [ \Pi_i(\tau,\s),\Pi_j(\tau,\s')] \,, \nonumber
\eea
leads, after some algebra, to 
\bea
\left[\xi_p^{(-n)},\xi_q^{(n)}    \right] &=& \e_p \,\, \delta_{pq} \,\, ( 64 n^2 \kappa^2 \mu^2)^{-1} {\w_n^- \over \omega_n^+ (\omega_n^+ - \kappa \mu)}
\sac p,q=1,2,5,6  \nn
\left[ \xi_p^{(-n)} ,  \xi_q^{(n)}\right] &=& \e_p \,\, \delta_{pq} \,\,( 64 n^2 \kappa^2 \mu^2)^{-1} {\w_n^+ \over \omega_n^- (\omega_n^+ - \kappa \mu)}
\sac p,q=3,4,7,8
\eea
where $\e_i$  are signs to be chosen positive for the modes associated to positive frequencies ($p=1,2,3,4$) and
negative for those associated to negative frequencies ($p=5,6,7,8$).

The rescalings needed to define standard normalized
creation and annihilation operators $[a^{(n)}_p,a^{(n) \dagger}_q]=\delta_{pq}$
are now obvious. A straightforward, though lengthy, computation
shows that the Hamiltonian becomes
\bea
H_{lc} &=& H_0 + \sum_{n>0}^{\infty} H_n \,,\nn
H_n &=& \omega_{n}^+ \left( N_1^{(n)} +N_2^{(n)} +N_5^{(n)} +N_6^{(n)} \right)
+\omega_{n}^- \left( N_3^{(n)} +N_4^{(n)} + N_7^{(n)} +N_8^{(n)} \right) \,.
\label{ham-bos}
\eea
Here $H_0$ is the zero-mode contribution, to be computed below, and $N^{(n)}_p=a^{(n) \dagger}_p a^{(n)}_p$
is the number operator for the corresponding modes. Note that, although generally taken
for granted, it was not obvious a priory that the coefficient multiplying each number operator would be precisely
its corresponding frequency; this allows us to think of the frequencies of the modes
as exactly the quanta of energy needed to create them.

To end up with the quantization of the string modes in the bosonic sector, we need to impose the level-matching condition
following from invariance under translations along the worldsheet of the closed string,
\be
\int_0^{\pi} \Pi_i X'^i = 0 \,.
\ee
After some algebra this translates into
\be
\sum_{n>0} \, n \, \sum_{i=1}^8 \, \e_i \, N^{(n)}_i \, = \, 0 \,,
\ee
which implies
\be \label{level-matching}
N_1^{(n)} +N_2^{(n)} +N_3^{(n)} +N_4^{(n)}
\,\, = \,\, N_5^{(n)} +N_6^{(n)} + N_7^{(n)} +N_8^{(n)} \sac n \neq 0 \,.
\ee
As in flat space, we need to add as many right-moving excitations to the ground state as
left-moving ones.

\subsubsection{The particle-like modes}
\label{sec-quant-2}

We now study the $n=0$ modes of the field~\bref{field-exp}.
A look at the frequencies~\bref{freqs}
shows that the extrapolation to $n=0$ leads to $\omega_0^+ = 2 \kappa \mu$ and $\omega_0^- = 0$.
For $\omega_0^+$ the story is much like in the $n\neq 0$ case. In particular, the eigenfunctions
are obtained by setting $n=0$ in the expressions~\bref{eigenfun}, and they lead to two zero-mode harmonic
oscillators.

However, the vanishing of $\omega_0^-$ leads one to suspect the appearance of a free particle spectrum.
Let us examine this case in detail. We look for the most general solution to the equations of motion~\bref{eom2} with
$n=0$. The first remark is that when $n=0$ the last term in~\bref{eom2} vanishes, so that,
the dynamics in the $y^1y^2$-plane  decouple from the dynamics in the $y^3y^4$-plane.
Let us concentrate on the first one, for which the equations of motion reduce to
\be
\label{eomzero}
\ddot{y}^i_n - 2 \kappa \,  f_{ij} \,\dot{y}^j_n  \, + \, \kappa^2 \, k_i y^i  =0 \sac i=1,2,
\ee
with
\be
f_{12}=-f_{21} = \mu G^{1/2} \sac (k_1,k_2)=(4\mu^2 \hg^2 G,0) \,.
\ee
Note that if we set the deformation to zero, then $k_i=0$,
so that the system reduces to a Landau problem (i.e. a charged particle moving in a plane threatened by
a constant magnetic field). It is well-known that the quantization of this system leads to a ground state with
an infinite (but discrete) degeneracy.\footnote{See~\cite{JO} for an analogous situation in a
certain Penrose limit of the $AdS_5\times T^{1,1}$ background and its dual $\caln=1$ SCFT.}

As we turn on the deformation, we slightly modify the value of the magnetic field and, more importantly, we introduce
a quadratic potential along one of the axis of the plane (in this case, along $y^1$). The problem is essentially that
of a massive charged particle moving in a sheet positively curved along $y^1$, as depicted in figure~\ref{fig},
and subject to gravity.
We would like to point out that our problem is formally identical to that of a particle moving in an anti-Mach metric.
See for instance~\cite{Blau:2003ia,Blau:2003rt}.
Indeed, it has been shown in these papers that some pp-wave solutions of type IIB supergravity {\it
with only the NS field turned on}
lead to an equation for the zero-modes which is similar to~\bref{eomzero} but
with the sign of the last term is reversed. This leads to a quadratic but repulsive interaction in the
Landau plane, i.e., it corresponds to bending the plane of figure \ref{fig} in a concave manner.
These authors showed that the infinite degeneracy of the Landau
ground state is completely broken, and a free-particle spectrum arises. However, because of their negative sign
of the quadratic interaction, the kinetic energy of such free-particles appears with the 'wrong' sign in the
hamiltonian, leading to a spectrum unbounded from below.

Our case is definitely better behaved. The reason why our supergravity equations are
satisfied for positive values of $k_{i}$  is due to the fact that we have turned on extra fluxes
other than the NS 3-form, namely the RR 3- and 5-forms. The ultimate consequence
will be that the free-particle spectrum will arise with the standard positive-definite kinetic term.

Let us proceed to the quantization of our problem. The general solution of~\bref{eomzero} is
\bea
y^1 &=& y^1_0  +  \sqrt{2 \a'} \, \left( [\xi_1^{(0)} \, u^1_1(\w_0^+) + \xi_2^{(0)} \,u^1_2(\w_0^+)] e^{i\w_0^+ \tau}
+ c.c. \right) \,, \nn
y^2 &=& y^2_0 + v_2 \, \tau  + \sqrt{2 \a'} \, \left( [\xi_1^{(0)} \, u^2_1(\w_0^+) + \xi_2^{(0)} \,u^2_2(\w_0^+)] e^{i\w_0^+ \tau}
+ c.c. \right) \,,
\label{zero-exp}
\eea
where $v_2 \equiv 2 \kappa \mu \hg^2 G^{1/2} y^1_0$ and, as mentioned above, $\w_0^+ = 2 \kappa \mu$ and
the values of the eigenfunctions $u^i_1$ and $u^i_2$ are given by setting $n=0$ in~\bref{eigenfun}.
The oscillator part describes the motion of a particle along (non-circular) orbits in the plane.
The constant and linear terms describe the position of the center of such orbits in the plane.
As we see, the center is allowed to move only along $y^2$, which is after all an isometry
of the plane, and of the metric in the original coordinates~\bref{ours}. The velocity $v_2$ is fixed by (and proportional to)
the position in the axis $y^1$ transverse to the collective motion.
Note however that the canonical momenta $(\pi_1,\pi_2)$ are non-zero in both directions because,
as follows from~\bref{can-mom},
\be \label{constraints}
\chi_1 \, \equiv \, \pi_1 \,+\, {\kappa \mu \over 2 \pi \a'}
 \, y^2_0 = 0 \sac \chi_2 \, \equiv \,  \pi_2 - {v_2 + \kappa \mu  y^1_0 \over 2 \pi \a'}\,   \,.
\ee
Recall however that the canonical momenta $\pi_i$ are gauge dependent.

Note that the relations~\bref{constraints} do not include the velocities, which means that they should
be regarded as phase space constraints. Another way of rephrasing it is that $\det {\pa L \over \pa \dot{y}^i \pa \dot{y}^j}=0$,
making it impossible to express the velocities in terms of the momenta. The Poisson bracket of the two constraints
is non-vanishing, implying that $(\chi_1,\chi_2)$ form a system of second-class constraints. The quantization procedure
is therefore straightforward: instead of promoting the Poisson bracket to a quantum commutator, one promotes the
Dirac bracket, defined for any pair of phase-space functions $F,G$ as
\be
\{ F , G \}_{D.B.} = \{  F , G \}_{P.B.} - \{  F , \chi_\a \}_{P.B.} C^{\a \b}  \{  \chi_\b , G \}_{P.B.} \,,
\ee
where $C_{\a\b} = \{  \chi_\a , \chi_\b \}_{P.B.}$, and $C^{\a\b}$ is its inverse. One of the nicest properties of
the Dirac bracket is that the constraints can be imposed either before of after taking the brackets, which implies that
we can 'solve' the relations~\bref{constraints} once and for all. We choose to solve for the $\pi_i$ in
terms of the $y^i_0$,
\be
\pi_1 \,=\, -{\kappa \mu \over 2 \pi \a'}
 \, y^2_0 \sac
\pi_2 \, = \,  {v_2 + \kappa \mu  y^1_0 \over 2 \pi \a'}\,   \,,
\ee
and so we are just left with the pair $(y^1_0,y^2_0)$ subject to the following Dirac bracket,
\be \label{noncomm}
\{ y^2_0 , y^1_0  \}_{D.B.} = { \a' \over \k \mu} \sqrt{1+\hg^2} \, = { 1 \over 2 \mu p_v } \sqrt{1+\hg^2} \,.
\ee
In other words, the position in the $y^1$-axis essentially becomes the momentum operator for the position
in the $y^2$-axis.
The standard Heinsenberg uncertainty now implies that the it is not possible
to resolve the system at distances smaller than the square root of the RHS of~\bref{noncomm}.
After an appropriate rescaling, we obtain the standard position-momentum commutation relations,
\be
p_{y^2} \equiv {2 \mu p_v \over \sqrt{1+\hg^2}}\, y^1_0 \espai \Rightarrow \espai \{y^2_0,p_{y^2}\}_{D.B.} =1 \,.
\ee

Adding the copy of this non-commutative system corresponding to the $(y^3,y^4)$-plane and
the two harmonic oscillators in~\bref{zero-exp}, the complete $n=0$ Hamiltonian reads
\be \label{zero-ham}
H_0 = {\kappa^2 \hg^2  \over 4 \a' p_v^2 G}  \, \left[(p_{y^2})^2 + (p_{y^4})^2 \right]
+ \omega^+_0 \left[ N_1^{(0)} + N_2^{(0)} \right] \,.
\ee

Let us conclude this section by commenting on the zero deformation limit. An obvious
result is that setting $\gamma=0$ removes the contribution to the energy
of the 'free-particle' excitations, leaving behind a highly degenerate ground state.
A not so immediate consequence is that this degeneracy becomes discrete as $\gamma\rightarrow 0$.
The reason is that the behavior of the differential system~\bref{eomzero} is not smooth
in such limit: for $\gamma=0$ all solutions to~\bref{eomzero} are of a harmonic oscillator type.
So when $\gamma=0$ we recover the well known infinite but discrete ground state degeneracy
of the Landau problem.
The non-smoothness of the limit can be understood from the fact that, no matter how small $\g$ is,
the extra term in the potential added to the Landau problem  radically changes the large $y^1$
asymptotics.

\subsection{Quantization of the fermionic sector}
\label{sec-quant-ferm}

The generalization of the flat space fermionic GS action to backgrounds with
null Killing vectors is~\cite{Metsaev:2001bj,Metsaev:2002re}
\bea \label{fer-lag}
S= {1 \over 2 \pi \a'} \int d\tau d\s \, \pa_\a X^M  \, \bar{\Theta} (\sqrt{-h} h^{\a\b} -\e^{\a\b} \tau_3 ) \, \tau_2  \,\G_M D_{\b} \Theta \,.
\eea
All conventions are explained in the appendix. Here we just remark that $\Theta$ is a pair of positive chirality
10d Majorana-Weyl spinors $\Theta^I$ $(I=1,2)$ which are rotated by the action of the $2\times 2$ Pauli matrices $(\tau_i)_{IJ}$.
The derivative $D_\a$ is the pull-back to the worldsheet of the type IIB supercovariant derivative.

Because the two 10d spinors $\Theta^I$ are Majorana-Weyl, we choose the following adapted basis for the $32\times 32$
gamma matrices
\be
\G^M =
\pmatrix{ 0 &  \gamma^M  \cr \bg^M  & 0 }
\sac \g^{M} \bg^{N} + \g^{N} \bg^{M} = 2 \eta^{MN} \,,
\ee
where $\gamma^M$ are the $16 \times 16$ matrices~\bref{sixteen}.
These $\gamma$-matrices should not be confused with $\hg$, the deformation parameter.
In this basis, positive chirality MW spinors can be simply written
\be
\Theta^I = \pmatrix{ \t^I  \cr 0 } \,,
\ee
with $\t^I$ a pair of real 16 component spinor.

Working in the light-cone gauge $\G_v \Theta^I=0$, or equivalently, $\bg^u \t^I=0$, the fermionic Lagrangian
~\bref{fer-lag} in our background~\bref{ours}-\bref{simpler} receives different types of contributions which we classify
according to their origin:
\bea
(2\pi \a') \, L_{kinetic}&=& 2i \kappa \,\, ( \t^1 \bg^- \pa_+ \t^1 + \t^2 \bg^- \pa_- \t^2  )  \,, \nn
(2\pi \a')L_{spin-connection}&=& -i {\kappa^2 \over 4}  \,\, (\t^1 \bg^- \sla{f}  \t^1 + \t^2 \bg^- \sla{f}  \t^2  )  \,, \nn
(2\pi \a')L_{H_3}&=& i {\kappa^2  \over 4}  \,\, (\t^1 \bg^- \sla{h}  \t^1 - \t^2 \bg^- \sla{h} \t^2  )  \,, \nn
(2\pi \a')L_{F_3}&=& -2 i \kappa^2 \, \hg \mu G^{1/2}  \,\, \t^1 \bg^- \g^{13}  \t^2   \,, \nn
(2\pi \a')L_{F_5}&=& 2 i \mu \kappa^2 G^{1/2}\,   \,\,\t^1 \bg^- \g^{1234}  \t^2     \,,
\eea
where we used the standard slash-notation (e.g. $\sla{f}=f_{ij} \g^{ij}$).
Note that whereas the first three terms
are chiral, those coming from the RR fields give rise to non-chiral 'mass terms'.

We now want to write down the equations of motion and solve them. It turns out to be convenient to choose
a specific representation for the $\gamma$-matrices which will allow us to solve the light-cone constraint and
reduce the $16 \times 16$ system into an $8\times 8$ one. Writing
\be
\g^i = \pmatrix{ 0 &  \rho^i  \cr (\rho^i)^T & 0 } \sac i=1,...,8\,,
\ee
and using the explicit representation~\bref{rhomatrices} for the $8 \times 8$ $\rho$-matrices,
we see that the light-cone constraint $\bg^u \t^I =0$ is automatically satisfied by spinors of the form
\be \label{minimal-spinor}
\theta^I = \pmatrix{ S^I  \cr 0 } \,,
\ee
with $S^I$ a pair of real but otherwise unrestricted 8-component spinors. The equations of motion in terms of
such spinors are rather simple-looking
\bea
-2\pa_+ S^1 + \frac{\kappa}{4} (\sla{f} -\sla{h}) S^1 + \mu G^{1/2} \kappa (\hg \rho^{13}-\rho^{1234}) S^2 &=&0 \,,\nn
-2\pa_- S^2 + \frac{\kappa}{4} (\sla{f} +\sla{h}) S^2 + \mu G^{1/2} \kappa (\hg \rho^{13}+\rho^{1234}) S^1 &=&0 \,,
\eea
where now slashed tensors imply contraction with the $\rho$ matrices.
Expanding in Fourier and frequency modes $S(\tau,\s)=\sum_{n} e^{2in \s} e^{i \w_n \tau}$, these equations become

\be
\pmatrix{ -i (\w_n + 2n)+{\kappa \over 4} (\sla{f} -\sla{h}) &  \mu G^{1/2} \kappa (\hg \rho^{13}-\rho^{1234})   \cr
\mu G^{1/2} \kappa (\hg \rho^{13}+\rho^{1234})                           & -i (\w_n - 2n)+{\kappa \over 4} (\sla{f} +\sla{h}) }
\pmatrix{ S^1 \cr S^2 } = 0 \,.
\ee
\vskip 0.3cm

\noindent We can now express $S^2$ as an algebraic function of $S^1$ and solve the resulting equations for $S^1$.
The requirement that the latter admit non-trivial solutions leads to
the restriction that the allowed frequencies are precisely the same as in the bosonic sector~\bref{freqs}.
This a consequence of having the supernumerary charges discussed in section~\ref{sec-susy}.

A further straightforward calculation shows that the Hamiltonian and the
level-matching constraints take exactly the same form as
the bosonic ones~\bref{ham-bos}~\bref{level-matching}, with the replacement of
bosonic creation/anihilation operators by fermionic ones.

A worth mentioning subtlety concerns the fermionic zero modes. On the one hand, we find 2 fermionic
zero-modes which are harmonic oscilators with $\w=2 \kappa \mu$: they are superpartners of the corresponding
bosonic ones. On the other hand, we find that the superpartners of the free-particle bosonic modes correspond
to two spinors constant on the worldsheet which do not contribute to the Hamiltonian, as expected from our supersymmetry
analysis in section \ref{sec-susy}.

\section{Field theory interpretation and discussion}
\label{sec-discussion}

Let us try to interpret the results of this paper, which were summarized in section \ref{sec-results}.
As far as the 4 directions coming from the $AdS_5$ is concerned, nothing
changes with respect to the maximally supersymmetric pp-wave~\cite{BMN}: exciting their modes corresponds
to the addition of spacetime derivatives to the operators.
We therefore discuss only the modes coming from the $S^5_\g$. Indeed, we will only
discuss the bosonic modes, as the extension to the fermionic ones is immediate.

We first concentrate on the particle-like ($n=0$) excitations of the string.
We found that there are two $n=0$
modes which contribute positively and with a continuous spectrum to the energy. Translating~\bref{zero-ham}
into field theory variables, we find that they contribute as
\be
\Delta -(J_{\phi_1} +J_{\phi_2}+ J_{\phi_3}) = \mbox{const.} \, \hg^2 \left[ (p_{y^2})^2 +(p_{y^4})^2  \right] \,,
\ee
where  $(p_{y^2},p_{y^4})$ refer to the momenta along the isometric directions of the two
modified Landau planes (see fig.\ref{fig}). To make contact between these two directions $(y^2,y^4)$
and the $U(1)_1\times U(1)_2$ along which the $SL(2,\CR)$ transformation was performed, recall
that the latter corresponded to shifts of $(\vp_1,\vp_2)$, and that in the Penrose limit we defined
(see eq.\ref{rescalings})
\be \label{anglerel}
\vp_1 = \left(1\over 2 G\right)^{1/2} {-y^2 +\sqrt{3} y^4 \over R} \sac
\vp_2 = -\left(1\over 2 G\right)^{1/2} {y^2 +\sqrt{3} y^4 \over R} \,.
\ee
A first comment is then that $(y^2,y^4)$ parametrise straight lines in the limit $R \rightarrow \infty$,
leading to a continuous spectrum for their respective momenta.
Now,~\bref{anglerel} induces the following relations: 
\be \label{p-and-J}
p_{y^2} = {1 \over R} \sqrt{1\over 2 G} \, \left(-J_{\vp_1} + J_{\vp_2}  \right) \sac
p_{y^4} = -{1 \over R} \sqrt{3  \over 2G} \, \left( J_{\vp_1}+J_{\vp_2} \right) \,.
\ee
Recall that, due to~\bref{angular}, $J_{\vp_1}$ and $J_{\vp_1}$ parametrize  how far
an operator like~\bref{half-bps} is from having $J_{\Phi_1}=J_{\Phi_2}=J_{\Phi_3}$.
Therefore, we conclude that {\it momentum along the isometric direction of the modified
Landau plane corresponds to charge under $U(1)_1\times U(1)_2$, which in turn
corresponds to departures from $J_{\Phi_1}=J_{\Phi_2}=J_{\Phi_3}$.}

This correspondence makes our results fit nicely with some properties which are well-known. On the one hand,
we know that only states which are charged under the $U(1)_1 \times U(1)_2$ suffer any modification
due to the $SL(2,\CR)$.\footnote{See~\cite{Gursoy:2005cn} for a recent application of this idea
to try to cure some of the unwanted features of the Maldacena-N\'u\~nez background.} The unique bosonic
vacuum that we obtain is precisely the single state out of the infinitely many of the original undeformed
Landau plane which is $U(1)_1\times U(1)_2$ invariant. It has $p_{y^2}=p_{y^4}=0$, and therefore
corresponds to the only state with $J_{\Phi_1}=J_{\Phi_2}=J_{\Phi_3}$.

Let us ask what happens to the other states in the undeformed Landau vacuum after the $SL(2,\CR)$
duality. At the classical level, we know that if a point-like state has momenta along the torus,
then the $SL(2,\CR)$ will map it to a state in the deformed background with both momenta and winding along the
torus~\cite{LM}. For certain values of $\g$, the final state can be interpreted as an extended spinning
string similar to those in~\cite{Frolov:2003qc}. Classical extended strings in the $\g$-deformation
of \ads were already analyzed by Lunin and Maldacena~\cite{LM}, and it is easy to see that they
are all decoupled in our Penrose limit.\footnote{This is due to the fact that, using the
language of section \ref{sec-penrose}, the extended
strings of~\cite{LM} are placed in the $S^5_\g$ in such a way that $\mu_i^2-\mu_j^2$
remains finite in the limit that we are taking. However, the region that we focus on
has $\mu_i^2-\mu_j^2 \sim 1/R$.} Nonetheless, some qualitative features can be imported
from this classical picture. In particular, the a priori point-like states of the
string in our background seem to have a 'minimum length' due to the non-commutative nature
of the quantum commutation relations~\bref{noncomm}.

Let us now try to explain why the discrete infinite degeneracy of
the Landau problem turns into a {\it gapless continuous} spectrum.
Equation~\bref{p-and-J} tells us that despite the fact that all $J_{\Phi_i}$
are of order $R^2$ in the Penrose limit, their differences must be kept of order $R$ in order
to yield finite worldsheet charges $p_y$, and thus contribute to the light-cone energy.
In other words, varying slightly the charges $p_y$
corresponds to roughly exchanging $R \sim N^{1/4}$ times the fields $\Phi_i$ in our operators.
In the large $N$ limit, the gap between the light-cone energy of operators which differ by a small number
of exchanges of the $\Phi_i$ operators tends to zero, and the spectrum becomes continuous.

The explicit prediction that we are doing here is that the conformal dimensions of these operators
in the deformed theory is given by
\be \label{main-formula}
\Delta-(J_{\phi_1} +J_{\phi_2}+ J_{\phi_3}) = g_{YM}^2 N \, \, {\g^2 \over 3} \,\, {J_{\vp_1}^2 + J_{\vp_1}^2+J_{\vp_1} J_{\vp_2}
\over J_{\phi_1} +J_{\phi_2}+ J_{\phi_3}} \,,
\ee
which follows from~\bref{zero-ham}, together with~\bref{string-charges},~\bref{rela} and~\bref{p-and-J},
and we recall that the definition of the $J_{\vp}$'s was given in~\bref{angular}.
Note that, in the Penrose limit, all quantities in the right hand side of~\bref{main-formula} scale
in such a way that the net result is finite.

\vskip .1cm

Having explained the two $n=0$ modes with continuous spectrum,
we turn to the two harmonic oscillator ones, which have $\Delta -(J_{\phi_1} +J_{\phi_2}+ J_{\phi_3})=2$.
The minimal choice is to assume that they correspond to insertions
of either $\bar{\Phi}_1$ or $\Phi_2 \bar{\Phi}_3$, since both have the correct $\Delta-\sum J$,
and the first has $p_{y^2}=0$ whereas the second has $p_{y^4}=0$.

Finally, the same reasoning as in the maximally supersymmetric pp-wave~\cite{BMN}
works here:
the stringy modes correspond to the same replacements/insertions mentioned above
with the addition of the corresponding momentum phases. In terms of field theory
variables, the contribution of the level $n$ modes to the anomalous dimensions is
\be
\Delta-(J_{\phi_1} +J_{\phi_2}+ J_{\phi_3}) = \pm 1 \, +\, \sqrt{1+  g^2_{YM} N \,
{n^2 \over \left(J_{\phi_1} +J_{\phi_2}+ J_{\phi_3}\right)^2} } \,,
\ee
which follows from~\bref{freqs},~\bref{string-charges}, and~\bref{rela}.

\vskip .3cm

An obvious interesting continuation of this work would be to check these predictions
in the field theory. This might be more difficult than in the original BMN case~\cite{BMN,Santambrogio:2002sb}
because of the large number of different phases that the deformation introduces
to operators with a large number of fields. It would also be nice to see if
similar phenomena to those studied here happen to $\g$-deformations of other SCFTs whose Penrose
limits lead to ground states with infinite discrete degeneracy. This is
the case, for example, of the Klebanov-Witten theory~\cite{Klebanov:1998hh}, which
has an $AdS_5 \times T^{1,1}$ dual admitting a Penrose limit which also leads to a Landau problem~\cite{JO,PandoZayas:2002rx}.

\bigskip\bigskip

\bigskip\noindent{\bf Note added:}
The same day that this paper was sent to the archive, the paper~\cite{deMelloKoch:2005vq} appeared
which has some overlap with sections~\ref{sec-penrose} and~\ref{sec-quant-1}.

\bigskip\bigskip

\bigskip\noindent{\bf Acknowledgements}
\par
\noindent I would like to thank Niklas Beisert,
Jerome Gauntlett, Sangmin Lee, Bogdan Stenfanski and Daniel Waldram
for many useful discussions. I would also like to thank the Perimeter Institute,
where part of this work was completed, as well as
the organizers of the {\it Workshop on Gravitational Aspects of String Theory} (May
2-6, 2005) at Fields Institute, Toronto, for their support and hospitality.

\bigskip\bigskip

\appendix

\section*{Appendix}

\section{Conventions}
\label{sec-appendix-conv}

In the background after the Penrose limit we have been consistent in the use of the
following notation

\begin{center}
\otaula{llcl}
curved spacetime coordinates: & $X^M$  & &$M,N = u,v,1,2,...8$  \,,\cr
Landau directions:         & $y^i$ & &$i,j=1,2,3,4$ \,,\cr
unchanged directions:       & $r^a$ && $a,b=5,6,7,8$ \,,\cr
worldsheet coordinates:       & $\s^\a$ && $(\tau,\sigma)$ \,,\cr
creation/anihilation modes:  & $\xi_p$ && $p,q=1,...,8$ \,.\cr
\ctaula
\end{center}
In the quantization of the fermionic part of the action, we defined the antisymmetric symbol $\e^{\a\b}$ such that
$\e^{01}=-1$. The derivative $D_{\a}$ appearing in~\bref{fer-lag} is the pull-back to the worldsheet
of the type IIB super covariant derivative,
\small
\bea
D_{M}&=&\nabla_M+ {1\over 8} \tau_3(H_3)_{MNP}  \G^{NP} -{ e^{\phi} \over 48}
\left[\tau_1 (F_3)_{NPQ}\G^{NPQ} + {i \over 40} \tau_2(F_5)_{NPQRS}\G^{NPQRS} \right] \G_M \,,\nn
\nabla_M &=& \partial_M + {1 \over 4} \w_M^{NP} \G_{NP} \nonumber \,.
\eea
\normalsize
In the process of reducing the dimensionality of the originally 32-component spinors,
we first used a Weyl representation
\be
\G^M =
\pmatrix{ 0 &  \gamma^M  \cr \bg^M  & 0 }
\sac \g^{\mu} \bg^{\nu} + \g^{\nu} \bg^{\mu} = 2 \eta^{\mu \nu} \,,
\ee
with $\G^M$ and $\g^M$ being $32 \times 32$ and $16 \times 16$ respectively, and
\be \label{sixteen}
\g^M=(1,\g^I,\g^9) \sac \bg^M=(-1,\g^I,\g^9) \sac I=1,...,8 \,.
\ee
We also used the convenient notation
\be
\g^{ij}=\undos (\g^i \bg^j-\g^j \bg^i) \sac \g^{ijk}={1 \over 6} (\g^i \bg^j \g^k + 5 \mbox{ terms} )
\sac \mbox{etc.}
\ee
Choosing the Majorana representation for the $\G$-matrices, all $\g$-matrices are real and symmetric.
Note that the 8 matrices $\g^I$ form a representation of $SO(8)$ Clifford algebra. We can use
the explicit representation for them in which (see e.g.~\cite{GSW})
\be
\g^I = \pmatrix{ 0 &  \rho^I  \cr (\rho^I)^T & 0 } \,,
\label{rhomatrices}
\ee
with
\begin{center}
\otaula{lcl}
$\rho^1 = \ex     \times \ex       \times \ex$ , & $\espai$ &   $\rho^2=1     \times \tau_1 \times \ex$ , \cr
$\rho^3 = 1      \times \tau_3   \times \ex  $, & &    $\rho^4=\tau_1\times \ex     \times 1 $ , \cr
$\rho^5 = \tau_3 \times \ex       \times 1 $,&    &     $\rho^6=\ex    \times 1      \times \tau_1 $, \cr
$\rho^7 = \ex     \times 1        \times \tau_3 $ ,& &  $\rho^8=1     \times 1      \times 1 $.
\ctaula
\end{center}
The light-cone gauge implies that $\g$-matrices will only end up acting on spinors of the form
~\bref{minimal-spinor}, so we can everywhere replace products of $\g$-matrices by products
of $\rho$-matrices.
\be
\g^{I_1} \bg^{I_2} ... \,\, \longrightarrow \,\, \rho^{I_1} (\rho^{I_2})^T ... \,.
\ee

\section{The full $\g$-deformation of \ads}
\label{sec-appendix-full}

For the sake of clarity of the exposition, in the paper
we avoided writing the full type IIB configuration corresponding to the $\g$-deformation
of \ads before the Penrose limit. Adapting to our notation the results of~\cite{LM} we have,
\bea \label{allofit}
ds^2_{\g} &=& R^2 \left[ ds^2_{AdS_5}+ \sum_{i=1}^3 \left( d\mu_i^2 + G \mu_i^2 d\phi_i^2 \right)
+R^4 \g^2 G \mu_1^2\mu_2^2\mu_3^2 (\sum_{i=1}^3 d\phi_i )^2 \right] \,, \nn
e^{2\phi}&=& e^{2\phi_0} G \,,\nn
B_2^{NS} &=& \g R^4 G \, ( \mu_1^2\mu_1^2 d\phi_1 d\phi_2 +\mu_2^2\mu_3^2 d\phi_2 d\phi_3 +\mu_3^2\mu_1^2 d\phi_3 d\phi_1 ) \,,\nn
C_2^{RR} &=& -12 \g R^4 e^{-\phi_0} \mu_1 \mu_2 \mu_3 \sin\a \,  d\a d\t d\psi \,, \nn
F_5^{RR} &=& 4 R^4 e^{-\phi_0} ( vol_{AdS_5} + G vol_{S^5} ) \,,
\eea
with
\bea
G^{-1} &\equiv& 1+R^4 \g^2 (\mu_1^2 \mu_2^2 + \mu_2^2 \mu_3^2 +\mu_1^2 \mu_3^2 ) \,, \nn
(\mu_1,\mu_2,\mu_3) &=& (\cos \a,\sin\a \cos\t,\sin\a \sin\t) \,.
\eea


\bibliographystyle{tonim}
\bibliography{Finalbib}

\providecommand{\href}[2]{#2}\begingroup\begin{thebibliography}{10}

\bibitem{Leigh:1995ep}
R.~G. Leigh and M.~J. Strassler, \emph{Exactly marginal operators and duality
  in four-dimensional N=1 supersymmetric gauge theory}, Nucl. Phys. {\bf B447}
  (1995) 95--136,
\href{http://www.arXiv.org/abs/hep-th/9503121}{{\tt hep-th/9503121}}

\bibitem{M}
J.~M. Maldacena, \emph{The large N limit of superconformal field theories and
  supergravity}, Adv. Theor. Math. Phys. {\bf 2} (1998) 231--252,
\href{http://www.arXiv.org/abs/hep-th/9711200}{{\tt hep-th/9711200}}

\bibitem{W}
E.~Witten, \emph{Anti-de Sitter space and holography}, Adv. Theor. Math. Phys.
  {\bf 2} (1998) 253--291,
\href{http://www.arXiv.org/abs/hep-th/9802150}{{\tt hep-th/9802150}}

\bibitem{GKP}
S.~S. Gubser, I.~R. Klebanov  and A.~M. Polyakov, \emph{Gauge theory
  correlators from non-critical string theory}, Phys. Lett. {\bf B428} (1998)
  105--114,
\href{http://www.arXiv.org/abs/hep-th/9802109}{{\tt hep-th/9802109}}

\bibitem{LM}
O.~Lunin and J.~Maldacena, \emph{Deforming field theories with U(1) x U(1)
  global symmetry and their gravity duals},
\href{http://www.arXiv.org/abs/hep-th/0502086}{{\tt hep-th/0502086}}

\bibitem{MR}
J.~M. Maldacena and J.~G. Russo, \emph{Large N limit of non-commutative gauge
  theories}, JHEP {\bf 09} (1999) 025,
\href{http://www.arXiv.org/abs/hep-th/9908134}{{\tt hep-th/9908134}}

\bibitem{Hashimoto:1999ut}
A.~Hashimoto and N.~Itzhaki, \emph{Non-commutative Yang-Mills and the AdS/CFT
  correspondence}, Phys. Lett. {\bf B465} (1999) 142--147,
\href{http://www.arXiv.org/abs/hep-th/9907166}{{\tt hep-th/9907166}}

\bibitem{Frolov:2005ty}
S.~A. Frolov, R.~Roiban  and A.~A. Tseytlin, \emph{Gauge - string duality for
  superconformal deformations of N = 4 super Yang-Mills theory},
\href{http://www.arXiv.org/abs/hep-th/0503192}{{\tt hep-th/0503192}}

\bibitem{Frolov:2005dj}
S.~Frolov, \emph{Lax pair for strings in Lunin-Maldacena background},
\href{http://www.arXiv.org/abs/hep-th/0503201}{{\tt hep-th/0503201}}

\bibitem{Beisert:2005if}
N.~Beisert and R.~Roiban, \emph{Beauty and the Twist: The Bethe Ansatz for
  Twisted N=4 SYM},
\href{http://www.arXiv.org/abs/hep-th/0505187}{{\tt hep-th/0505187}}

\bibitem{Benvenuti:2005cz}
S.~Benvenuti and M.~Kruczenski, \emph{Semiclassical strings in Sasaki-Einstein
  manifolds and long operators in N = 1 gauge theories},
\href{http://www.arXiv.org/abs/hep-th/0505046}{{\tt hep-th/0505046}}

\bibitem{Benvenuti:2005ja}
S.~Benvenuti and M.~Kruczenski, \emph{From Sasaki-Einstein spaces to quivers
  via BPS geodesics: Lpqr},
\href{http://www.arXiv.org/abs/hep-th/0505206}{{\tt hep-th/0505206}}

\bibitem{Ahn:2005vc}
C.~Ahn and J.~F. Vazquez-Poritz, \emph{Marginal deformations with U(1)**3
  global symmetry},
\href{http://www.arXiv.org/abs/hep-th/0505168}{{\tt hep-th/0505168}}

\bibitem{Gauntlett:2005jb}
J.~P. Gauntlett, S.~Lee, T.~Mateos  and D.~Waldram, \emph{Marginal Deformations
  of Field Theories with $AdS_4$ Duals},
\href{http://www.arXiv.org/abs/hep-th/0505207}{{\tt hep-th/0505207}}

\bibitem{Berenstein:2000hy}
D.~Berenstein and R.~G. Leigh, \emph{Discrete torsion, AdS/CFT and duality},
  JHEP {\bf 01} (2000) 038,
\href{http://www.arXiv.org/abs/hep-th/0001055}{{\tt hep-th/0001055}}

\bibitem{Berenstein:2000ux}
D.~Berenstein, V.~Jejjala  and R.~G. Leigh, \emph{Marginal and relevant
  deformations of N = 4 field theories and non-commutative moduli spaces of
  vacua}, Nucl. Phys. {\bf B589} (2000) 196--248,
\href{http://www.arXiv.org/abs/hep-th/0005087}{{\tt hep-th/0005087}}

\bibitem{NP}
V.~Niarchos and N.~Prezas, \emph{BMN operators for N = 1 superconformal
  Yang-Mills theories and associated string backgrounds}, JHEP {\bf 06} (2003)
  015,
\href{http://www.arXiv.org/abs/hep-th/0212111}{{\tt hep-th/0212111}}

\bibitem{BMN}
D.~Berenstein, J.~M. Maldacena  and H.~Nastase, \emph{Strings in flat space and
  pp waves from N = 4 super Yang Mills}, JHEP {\bf 04} (2002) 013,
\href{http://www.arXiv.org/abs/hep-th/0202021}{{\tt hep-th/0202021}}

\bibitem{Cvetic:2002nh}
M.~Cvetic, H.~Lu, C.~N. Pope  and K.~S. Stelle, \emph{Linearly-realised
  worldsheet supersymmetry in pp-wave background}, Nucl. Phys. {\bf B662}
  (2003) 89--119,
\href{http://www.arXiv.org/abs/hep-th/0209193}{{\tt hep-th/0209193}}

\bibitem{JO}
J.~Gomis and H.~Ooguri, \emph{Penrose limit of N = 1 gauge theories}, Nucl.
  Phys. {\bf B635} (2002) 106--126,
\href{http://www.arXiv.org/abs/hep-th/0202157}{{\tt hep-th/0202157}}

\bibitem{Metsaev:2001bj}
R.~R. Metsaev, \emph{Type IIB Green-Schwarz superstring in plane wave Ramond-
  Ramond background}, Nucl. Phys. {\bf B625} (2002) 70--96,
\href{http://www.arXiv.org/abs/hep-th/0112044}{{\tt hep-th/0112044}}

\bibitem{Metsaev:2002re}
R.~R. Metsaev and A.~A. Tseytlin, \emph{Exactly solvable model of superstring
  in plane wave Ramond- Ramond background}, Phys. Rev. {\bf D65} (2002) 126004,
\href{http://www.arXiv.org/abs/hep-th/0202109}{{\tt hep-th/0202109}}

\bibitem{Russo:2002qj}
J.~G. Russo and A.~A. Tseytlin, \emph{A class of exact pp-wave string models
  with interacting light-cone gauge actions}, JHEP {\bf 09} (2002) 035,
\href{http://www.arXiv.org/abs/hep-th/0208114}{{\tt hep-th/0208114}}

\bibitem{Mizoguchi:2002qy}
S.~Mizoguchi, T.~Mogami  and Y.~Satoh, \emph{Penrose limits and Green-Schwarz
  strings}, Class. Quant. Grav. {\bf 20} (2003) 1489--1502,
\href{http://www.arXiv.org/abs/hep-th/0209043}{{\tt hep-th/0209043}}

\bibitem{Green:1983wt}
M.~B. Green and J.~H. Schwarz, \emph{Covariant description of superstrings},
  Phys. Lett. {\bf B136} (1984)
367--370

\bibitem{Blau:2003rt}
M.~Blau, M.~O'Loughlin, G.~Papadopoulos  and A.~A. Tseytlin, \emph{Solvable
  models of strings in homogeneous plane wave backgrounds}, Nucl. Phys. {\bf
  B673} (2003) 57--97,
\href{http://www.arXiv.org/abs/hep-th/0304198}{{\tt hep-th/0304198}}

\bibitem{Blau:2001ne}
M.~Blau, J.~Figueroa-O'Farrill, C.~Hull  and G.~Papadopoulos, \emph{A new
  maximally supersymmetric background of IIB superstring theory}, JHEP {\bf 01}
  (2002) 047,
\href{http://www.arXiv.org/abs/hep-th/0110242}{{\tt hep-th/0110242}}

\bibitem{Blau:2003ia}
M.~Blau, P.~Meessen  and M.~O'Loughlin, \emph{Goedel, Penrose, anti-Mach: Extra
  supersymmetries of time- dependent plane waves}, JHEP {\bf 09} (2003) 072,
\href{http://www.arXiv.org/abs/hep-th/0306161}{{\tt hep-th/0306161}}

\bibitem{Gursoy:2005cn}
U.~Gursoy and C.~Nunez, \emph{Dipole Deformations of N=1 SYM and Supergravity
  backgrounds with U(1) X U(1) global symmetry},
\href{http://www.arXiv.org/abs/hep-th/0505100}{{\tt hep-th/0505100}}

\bibitem{Frolov:2003qc}
S.~Frolov and A.~A. Tseytlin, \emph{Multi-spin string solutions in AdS(5) x
  S**5}, Nucl. Phys. {\bf B668} (2003) 77--110,
\href{http://www.arXiv.org/abs/hep-th/0304255}{{\tt hep-th/0304255}}

\bibitem{Santambrogio:2002sb}
A.~Santambrogio and D.~Zanon, \emph{Exact anomalous dimensions of N = 4
  Yang-Mills operators with large R charge}, Phys. Lett. {\bf B545} (2002)
  425--429,
\href{http://www.arXiv.org/abs/hep-th/0206079}{{\tt hep-th/0206079}}

\bibitem{Klebanov:1998hh}
I.~R. Klebanov and E.~Witten, \emph{Superconformal field theory on threebranes
  at a Calabi-Yau singularity}, Nucl. Phys. {\bf B536} (1998) 199--218,
\href{http://www.arXiv.org/abs/hep-th/9807080}{{\tt hep-th/9807080}}

\bibitem{PandoZayas:2002rx}
L.~A. Pando~Zayas and J.~Sonnenschein, \emph{On Penrose limits and gauge
  theories}, JHEP {\bf 05} (2002) 010,
\href{http://www.arXiv.org/abs/hep-th/0202186}{{\tt hep-th/0202186}}

\bibitem{deMelloKoch:2005vq}
R.~de~Mello~Koch, J.~Murugan, J.~Smolic  and M.~Smolic, \emph{Deformed PP-waves
  from the Lunin-Maldacena Background},
\href{http://www.arXiv.org/abs/hep-th/0505227}{{\tt hep-th/0505227}}

\bibitem{GSW}
M.~B. Green, J.~H. Schwarz  and E.~Witten, \emph{Superstring Theory. Vol. 1},
Cambridge, Uk: Univ. Pr. ( 1987) 469 P. ( Cambridge Monographs On Mathematical
  Physics)

\end{thebibliography}\endgroup



\end{document}